\documentstyle[prd,aps,preprint,tighten,epsfig]{revtex}

\begin{document}

\draft

\title{Updated Values of Running Quark and Lepton Masses}
\author{{\bf Zhi-zhong Xing}
\thanks{E-mail: xingzz@mail.ihep.ac.cn},
~ {\bf He Zhang}
\thanks{E-mail: zhanghe@mail.ihep.ac.cn},
~ {\bf Shun Zhou}
\thanks{E-mail: zhoush@mail.ihep.ac.cn}}
\address{
Institute of High Energy Physics, Chinese Academy of Sciences,
Beijing 100049, China}

\maketitle

\begin{abstract}
Reliable values of quark and lepton masses are important for model
building at a fundamental energy scale, such as the Fermi scale
$M^{}_Z \approx 91.2$ GeV and the would-be GUT scale
$\Lambda^{}_{\rm GUT} \sim 2 \times 10^{16}$ GeV. Using the latest
data given by the Particle Data Group, we update the running quark
and charged-lepton masses at a number of interesting energy scales
below and above $M^{}_Z$. In particular, we take into account the
possible new physics scale ($\mu \sim 1$ TeV) to be explored by the
LHC and the typical seesaw scales ($\mu \sim 10^9$ GeV and $\mu \sim
10^{12}$ GeV) which might be relevant to the generation of neutrino
masses. For illustration, the running masses of three light Majorana
neutrinos are also calculated. Our up-to-date tables of running
fermion masses are expected to be very useful for the study of
flavor dynamics at various energy scales.
\end{abstract}


\section{Introduction}

In the standard model (SM) of electroweak interactions \cite{SM},
it is the Higgs mechanism that provides a self-consistent
framework to generate masses for gauge bosons and charged
fermions. Three neutrinos are exactly massless as a
straightforward consequence of the symmetry structure of the SM.
But this framework itself can neither predict the values of six
quark masses and three charged-lepton masses nor interpret the
observed hierarchy of their spectra \cite{FX00}. On the other
hand, current neutrino oscillation experiments
\cite{SNO,SK,KM,K2K} indicate that neutrinos are actually massive
and lepton flavors are mixed. Hence one has to extend the SM in
order to gain an insight into the dynamics of fermion mass
generation and flavor mixing. Possible new physics beyond the SM
is expected to appear far above the Fermi scale $M^{}_Z \approx
91.2$ GeV which essentially characterizes the scale of electroweak
symmetry breaking. For instance, the typical scale of grand
unified theories (GUTs) for strong and electroweak interactions
\cite{GUT} would be $\Lambda^{}_{\rm GUT} \sim 2 \times 10^{16}$
GeV, while $\mu \sim 1$ TeV is the possible scale of new physics
necessary to stabilize the mass of the Higgs boson and offer a
natural explanation of electroweak symmetry breaking \cite{HOOFT}.
Whenever a specific mass model of leptons and (or) quarks is built
at a certain high energy scale, one has to make use of the
technique of renormalization group equations (RGEs) \cite{RGE} to
bridge the gap between the model predictions (at $\mu \gg M^{}_Z$)
and the experimental data (at $\mu \lesssim M^{}_Z$). Therefore,
reliable values of running quark and lepton masses are very
important for building new physics models and testing their
viability.

A systematic estimate of running quark masses at various energy
scales was made by Fusaoka and Koide \cite{Koide98} in 1998. Their
numerical results have proved to be very useful for people working
on both hadronic physics at relatively low energies and electroweak
physics and GUTs at superhigh energies. However, a nontrivial update
of this work is greatly desirable today because a lot of changes in
particle physics have happened since 1998. On the one hand, some new
and more accurate data on fermion masses at low scales have been
accumulated \cite{PDG}. On the other hand, intensive interest has
been paid to some new physics scales such as $\mu \sim 1$ TeV to be
explored by the Large Hadron Collider (LHC) and $\mu \sim 10^{12}$
GeV which might be relevant to the seesaw mechanism of neutrino mass
generation \cite{SS}. These energy scales were not considered in
Ref. \cite{Koide98}, nor were the running masses of three light
neutrinos. Thus we are well motivated to do a new analysis of
running fermion masses at a variety of fundamental or interesting
energy scales.

Let us remark that the present paper is different from Ref.
\cite{Koide98} and other previous works in the following aspects:
\begin{itemize}
\item The running masses of three light Majorana neutrinos are calculated.
A global analysis of current neutrino oscillation data yields strong
constraints on two neutrino mass-squared differences ($\Delta
m^2_{21} \equiv m^2_2 - m^2_1 = (7.2 \cdots 8.9) \times 10^{-5}~{\rm
eV}^2$ and $|\Delta m^2_{32}| \equiv |m^2_3 - m^2_2| = (1.7 \cdots
3.3) \times 10^{-3}~{\rm eV}^2$) and three neutrino mixing angles
($30^\circ < \theta^{}_{12} < 38^\circ$, $36^\circ < \theta^{}_{23}
< 54^\circ$ and $\theta^{}_{13} < 10^\circ$) at the $99\%$
confidence level \cite{SV}. In addition, the latest cosmological
constraint on the absolute values of $m^{}_i$ is $m^{}_1 + m^{}_2 +
m^{}_3 < 0.61~{\rm eV}$ at the $95\%$ confidence level \cite{WMAP5}.
Since the unification of leptons and quarks is only possible at a
superhigh energy scale (e.g., the GUT scale or the seesaw scale), it
makes sense to calculate the running masses of three light neutrinos
together with those of charged fermions. We shall illustrate the
running effects of $m^{}_i (\mu)$ for $\mu \gg M^{}_Z$ by using the
RGEs both in the SM and in the minimal supersymmetric standard model
(MSSM).

\item The up-to-date values of quark masses \cite{PDG} given by the Particle
Data Group (PDG) at low energies are adopted, and a few new energy
scales (such as $\mu \sim 1$ TeV and $10^{12}$ GeV) are taken into
account. Any models for new physics beyond the SM may introduce
new energy scales above the Fermi scale, among which the $\mu \sim
1$ TeV scale is most appealing because it will soon be probed by
the LHC. Possible new physics at this energy frontier is likely to
be responsible for the origin of fermion masses and flavor mixing,
or it can at least shed light on these fundamental problems. On
the other hand, the intriguing seesaw \cite{SS} and leptogenesis
\cite{FY} mechanisms have motivated a lot of neutrino physicists
to study the properties of heavy Majorana neutrinos in order to
simultaneously account for the lightness of three known neutrinos
and the matter-antimatter asymmetry of our universe. The
particularly interesting mass scales of such heavy Majorana
particles are $\mu \sim 10^9$ GeV to $10^{12}$ GeV, where the
flavor effects on leptogenesis \cite{Flavor} have to be treated
carefully. Hence we shall make use of the RGEs to run the lepton
and quark masses to these new energy scales, just for the
convenience of model builders.

\item The treatment of quark masses crossing the flavor thresholds is
improved. We shall use the matching conditions only at the flavor
thresholds $\mu \equiv m^{}_Q(m^{}_Q)$ (for $Q = c, b, t$) and
calculate light quark masses at any other high energy scales with
the help of the RGEs. This approach has been clearly described in
Ref. \cite{Kniehl}. In comparison, the running quark masses
between two neighboring flavor thresholds were just computed with
the matching conditions in Ref. \cite{Koide98}. The
running-matching-running scheme proposed in Ref. \cite{Kniehl} and
used in our calculations is no doubt more reasonable.

\item The effective coupling constant $\alpha^{}_s$, which governs the strength of
strong interactions, is calculated by numerically solving its RGE.
In contrast, the values of $\alpha^{}_s$ at different energy scales
were evaluated via the analytical relation between $\alpha^{}_s$ and
the asymptotic scale parameter $\Lambda$ in Ref. \cite{Koide98}. The
latter method may result in some unnecessary uncertainties due to
the expansion of $1/\ln \left(\mu^2/\Lambda^2\right)$, as shown in
Refs. \cite{Kniehl,Lambda}.

\end{itemize}
Our main numerical results will be tabulated, as done in Ref.
\cite{Koide98}, to serve for a useful reference in building specific
models at various energy scales.

The remaining parts of this paper are organized as follows. In Sec.
II, we summarize the input data which include the current masses of
three light quarks at $\mu = 2~{\rm GeV}$, the values of
$m^{}_c(m^{}_c)$ and $m^{}_b(m^{}_b)$, the pole mass of the top
quark, the strong gauge coupling $\alpha^{}_s$ and the fine
structure constant $\alpha$ at $M^{}_Z$. We also collect the
relevant formulas for the RGEs and matching conditions of quark
masses and $\alpha^{}_s$. The strategy to deal with the flavor
thresholds and evolve the charged fermion masses to the Fermi scale
is outlined. Sec. III is devoted to the calculations of running
fermion masses up to the GUT scale both in the SM and in the MSSM.
Indeed, the true running parameters above the Fermi scale are the
Yukawa couplings of leptons and quarks because these particles can
only acquire their masses after the electroweak symmetry breaking
(i.e., below the Fermi scale). The running masses of three light
neutrinos are also illustrated for completeness. Finally, we make a
brief summary of our main results in Sec. IV.

\section{Below the Fermi Scale}

First of all, let us give some concise comments on the definition of
fermion masses. It is important to specify the theoretical framework
when discussing quark masses, since they are
renormalization-scheme-dependent. There are two very common
renormalization schemes. One of them is the on-shell scheme, in
which the position of the pole is the definition of the physical
mass $M$. The other is to define the {\it running} or renormalized
mass $m(\mu)$ in the dimensional regularization scheme with the
modified minimal subtraction ($\overline{\rm MS}$), where $\mu$
denotes the renormalization scale. Both definitions are suitable for
charged leptons. But the situation is quite different in the quark
sector due to the nonperturbative nature of the quantum
chromodynamics (QCD) at low energies: light quarks are always
confined in hadrons and can never be observed directly. Even for the
heaviest quark, the top quark, it is impossible to completely
eliminate the nonperturbative effect on the value of $m^{}_t$
extracted from the direct measurements. Hence the pole mass of a
quark is not well-defined. Note that the QCD Lagrangian has a chiral
symmetry in the limit where all quark masses are vanishing. The
scale of dynamical chiral symmetry breaking is $\Lambda^{}_\chi \sim
1~ {\rm GeV}$ \cite{Manohar}, which can be used to distinguish
between light ($m^{}_q < \Lambda^{}_\chi$ for $q= u, d, s$) and
heavy ($m^{}_q > \Lambda^{}_\chi$ for $q = c, b, t$) quarks.

Another important point is the decoupling of heavy flavors.
Because of the hierarchical mass spectrum of quarks, one should
integrate out the heavy degrees of freedom when considering the
properties of light flavors. Therefore, some consistent matching
conditions should be established between the full and effective
theories at the scales characterized by the masses of heavy
flavors. We shall work in the framework elaborated on in Ref.
\cite{Kniehl}.

\subsection{Running Quark Masses}

With the help of the chiral symmetry, one may extract the values of
$m^{}_u/m^{}_d$ and $m^{}_s/m^{}_d$ from the masses of pion and kaon
in a way independent of the renormalization scale \cite{GL}. The
mass of the strange quark can be determined from the spectral
function sum rules or lattice QCD simulations \cite{Manohar,GL}. The
up-to-date values of $m^{}_u$, $m^{}_d$ and $m^{}_s$ given by the
PDG are \cite{PDG}
\begin{eqnarray}
m^{}_u(2~{\rm GeV}) & = & 1.5 \sim 3.0  ~{\rm MeV} \ , \nonumber \\
m^{}_d(2~{\rm GeV}) & = & 3 \sim 7  ~{\rm MeV} \ ,  \\
m^{}_s(2~{\rm GeV}) & = & 95 \pm 25 ~{\rm MeV} \ . \nonumber
\end{eqnarray}
Note that these running masses are evaluated at $\mu = 2 ~ {\rm
GeV}$ with three active quark flavors ($u,d,s$). The evolution of
$m^{}_q(\mu)$ (for $q=u, d, s, c, b, t$) is governed by the
following RGE:
\begin{eqnarray}
\mu^2\frac{{\rm d} m^{}_q (\mu)}{{\rm d} \mu^2} =
-\gamma(\alpha^{}_s)m^{}_{q}(\mu) = - \sum_{i=0}^\infty \gamma^{}_i
\left(\frac{\alpha_s(\mu)}{\pi}\right)^{i+1} m^{}_q(\mu) \ ,
\end{eqnarray}
where $\alpha_s(\mu) \equiv g^2_s/(4\pi)$ is the effective coupling
constant of strong interactions with $g^{}_s$ being the strong gauge
coupling. The values of $\gamma^{}_i$ (for $i=0, 1, 2, 3$) are given
by \cite{dimension,4loop,c}
\begin{eqnarray}
\gamma^{}_0 & = & 1 \ , \nonumber \\
\gamma^{}_1 & = & \frac{1}{16}\left(\frac{202}{3}-\frac{20}{9}n^{}_q\right)  \ , \nonumber \\
\gamma^{}_2 & = & \frac{1}{64} \left[ 1249 - \left ( \frac{2216}{27}
+ \frac{160}{3}\zeta(3)\right) n^{}_q - \frac{140}{81} n^2_q \  \right], \nonumber \\
\gamma^{}_3 & = & \frac{1}{256}\left[
\frac{4603055}{162}+\frac{135680}{27}\zeta(3)
-8800\zeta(5) \right.\nonumber \\
&& -\left(
\frac{91723}{27}+\frac{34192}{9}-880\zeta(4)-\frac{18400}{9}\zeta(5)
\right)n^{}_q  \nonumber \\
&& \left. +
\left(\frac{5242}{243}+\frac{800}{9}\zeta(3)-\frac{160}{3}\zeta(4)\right)n^2_q
- \left(\frac{332}{243}-\frac{64}{27}\zeta(3)\right)n^2_q \right] \
,
\end{eqnarray}
where $n^{}_q$ is the number of active quark flavors with masses
$m^{}_q < \mu$; $\zeta(3) \approx 1.202057$, $\zeta(4) = \pi^4/90
\approx 1.082323$ and $\zeta(5) \approx 1.036928$ are the Riemann
zeta functions. To find the solution to Eq. (2), we write out the
detailed $\mu$-dependence of $\alpha_s(\mu)$ in terms of the
Callan-Symanzik beta function \cite{PDG},
\begin{eqnarray}
\mu^2 \frac{\partial \alpha^{}_s(\mu)}{\partial \mu^2} =
\beta(\alpha_s(\mu)) = - \sum_{i \geq 0} \beta^{}_i
\frac{\alpha^{i+2}_s (\mu)}{\pi^{i+1}}\ ,
\end{eqnarray}
where
\begin{eqnarray}
\beta^{}_0 & = &\frac{1}{4} \left( 11-\frac{2}{3}n^{}_q  \right )\ , \nonumber \\
\beta^{}_1 & = & \frac{1}{16} \left( 102-\frac{38}{3}n^{}_q \right) \ , \nonumber  \\
\beta^{}_2 & = & \frac{1}{64}\left(
\frac{2857}{2}-\frac{5033}{18}n^{}_q + \frac{325}{54}n^2_q \right) \
, \nonumber \\
\beta^{}_3 & = & \frac{1}{256} \left[\frac{149753}{6} + 3564\zeta(3)
- \left(\frac{1078361}{162} + \frac{6508}{27}\zeta(3)\right)
n^{}_q\right. \nonumber \\
&& + \left. \left(\frac{50065}{162} +
\frac{6472}{81}\zeta(3)\right)n^2_q + \frac{1093}{729}n^3_q \right]
\; .
\end{eqnarray}
Note that we adopt the $\overline{{\rm MS}}$ scheme for the RGEs
throughout this paper. The solution to Eq. (2) can be expressed as
\cite{c,pole}
\begin{equation}
m^{}_q(\mu) = {{\cal R}(\alpha^{}_s(\mu))} \hat{m}^{}_q \ ,
\end{equation}
where $\hat{m}^{}_q$ denotes the renormalization-invariant quark
mass, and
\begin{equation}
{\cal R}(\alpha^{}_s) =
\left(\frac{\alpha^{}_s}{\pi}\right)^{\gamma^{}_0/\beta^{}_0}
\left[1 + \frac{\alpha^{}_s}{\pi}{\cal C}^{}_1  +
\frac{\alpha^2_s}{2\pi^2} \left({\cal C}^2_1 + {\cal C}^{}_2\right)
+ \frac{\alpha^3_s}{\pi^3} \left( \frac{1}{6} {\cal C}^3_1 +
\frac{1}{2} {\cal C}^{}_1 {\cal C}^{}_2 + \frac{1}{3}{\cal C}^{}_3
\right) \right] \; .
\end{equation}
In Eq. (7), the terms of ${\cal O}(\alpha^4_s)$ and smaller have
been omitted and the coefficients ${\cal C}^{}_{1,2,3}$ are defined
as
\begin{eqnarray}
{\cal C}^{}_1 & = & \frac{\gamma^{}_1}{\beta^{}_0} -
\frac{\beta^{}_1 \gamma^{}_0}{\beta^2_0} \; , \nonumber \\
{\cal C}^{}_2 & = & \frac{\gamma^{}_2}{\beta^{}_0} -
\frac{\beta^{}_1 \gamma^{}_1}{\beta^2_0} - \frac{\beta^{}_2
\gamma^{}_0}{\beta^2_0} + \frac{\beta^2_1 \gamma^{}_0}{\beta^3_0} \
,
\nonumber \\
{\cal C}^{}_3 & = & \frac{\gamma^{}_3}{\beta^{}_0} -
\frac{\beta^{}_1\gamma^{}_2}{\beta^2_0} +
\frac{\beta^{2}_1\gamma^{}_1}{\beta^3_0} -
\frac{\beta^{}_2\gamma^{}_1}{\beta^2_0} -
\frac{\beta^{3}_1\gamma^{}_0}{\beta^4_0}
+2\frac{\beta^{}_1\beta^{}_2\gamma^{}_0}{\beta^3_0}
-\frac{\beta^{}_3\gamma^{}_0}{\beta^2_0} \ .
\end{eqnarray}
In a theory with multiple energy scales, such as the QCD with
$n^{}_l = n^{}_Q -1$ massless quarks and one heavy quark $Q$ (for
$m^{}_Q \gg \mu$), one should integrate out the heavy field and
construct an effective theory by requiring its consistency with the
full $n^{}_Q$-flavor theory at the heavy quark threshold
$\mu^{(n^{}_Q)} = {\cal O}(m^{}_Q)$. In this sense, Eq. (4) is valid
between two quark thresholds, which are defined by $\mu^{(n^{}_q)} =
m^{}_q (m^{}_q)$. Then we are in a position to address the
decoupling of heavy quarks and consider the matching conditions at
the flavor thresholds. In our calculations, we use the matching
relation between the strong coupling constants of the neighboring
flavors \cite{Kniehl}:
\begin{eqnarray}
\alpha^{(n^{}_q-1)}_s (\mu)& = & \zeta^2_g \alpha^{(n^{}_q)}_s(\mu)
\ ,
\end{eqnarray}
where $\zeta^2_g$ is already known at the three-loop level, i.e.,
\begin{eqnarray}
\zeta^2_g & = & 1 -  \frac{\alpha^{(n^{}_q)}_s(\mu)}{\pi}
\left(\frac{1}{6}\ln
\frac{\mu^2}{\left(\mu^{(n^{}_q)}\right)^2}\right) +\left(
\frac{\alpha^{(n^{}_q)}_s(\mu)}{\pi} \right)^2 \left(
\frac{11}{12}-\frac{11}{24}\ln
\frac{\mu^2}{\left(\mu^{(n^{}_q)}\right)^2} + \frac{1}{36}
\ln^2\frac{\mu^2}{\left(\mu^{(n^{}_q)}\right)^2} \right) \nonumber \\
& & + \left( \frac{\alpha^{(n^{}_q)}_s(\mu)}{\pi} \right)^3 \left[
\frac{564731}{124416}-\frac{82043}{27648}\zeta(3) -
\frac{955}{576}\ln \frac{\mu^2}{\left(\mu^{(n^{}_q)}\right)^2} +
\frac{53}{576}\ln^2
\frac{\mu^2}{\left(\mu^{(n^{}_q)}\right)^2} \right. \nonumber \\
&& \left. - \frac{1}{216}\ln^3
\frac{\mu^2}{\left(\mu^{(n^{}_q)}\right)^2} -
(n^{}_q-1)\left(\frac{2633}{31104}-\frac{67}{576}\ln
\frac{\mu^2}{\left(\mu^{(n^{}_q)}\right)^2} + \frac{1}{36}\ln^2
\frac{\mu^2}{\left(\mu^{(n^{}_q)}\right)^2}\right) \right] \ ,
\end{eqnarray}
in the $\overline{\rm MS}$ scheme.

Now we consider the input values of three heavy quark masses. For
charm and bottom quarks, their masses extracted from the heavy quark
effective theory, lattice gauge theory and QCD sum rules are
consistent with one another if they are all spelled out in the same
scheme and at the same scale \cite{PDG}:
\begin{eqnarray}
m^{}_c(m^{}_c) & = & 1.25 \pm 0.09 ~ {\rm GeV} \; ,
\nonumber \\
m^{}_b(m^{}_b) & = & 4.20 \pm 0.07 ~ {\rm GeV} \; .
\end{eqnarray}
The top quark can be directly measured because its lifetime is
shorter than the typical time scale of nonperturbative strong
interactions $\Lambda^{-1}_{\rm QCD}$ \cite{smith}, where the
magnitude of $\Lambda^{}_{\rm QCD}$ is about several hundred MeV.
The pole mass of the top quark extracted from the average of
several direct measurements is \cite{PDG}
\begin{equation}
M^{}_t = 172.5 \pm 2.7 ~{\rm GeV} \; .
\end{equation}
Note that the nonperturbative contribution to the top quark mass
may be proportional to $\Lambda^{}_{\rm QCD}$, which is much
smaller than the present experimental error bar. The pole mass
$M^{}_q$ (for $q=c, b, t$) can be translated into the running mass
at $\mu = M^{}_q$ through \cite{pole}
\begin{eqnarray}
M^{}_q= m^{}_q(M^{}_q) \left[1
+\frac{4}{3}\frac{\alpha^{}_s(M^{}_q)}{\pi}+ K^{(2)}_q
\left(\frac{\alpha^{}_s(M^{}_q)}{\pi}\right)^2 + K^{(3)}_q
\left(\frac{\alpha^{}_s(M^{}_q)}{\pi}\right)^3 \right]
\end{eqnarray}
to the accuracy of ${\cal O}(\alpha_s^3)$, where $\{K^{(2)}_c =
11.21$, $K^{(2)}_b = 10.17$, $K^{(2)}_t = 9.13\}$ and $\{K^{(3)}_c =
123.8$, $K^{(3)}_b = 101.5$, $K^{(3)}_t = 80.4\}$ are computed with
some typical values of the pole masses of light quarks \cite{pole}.
In addition to the matching condition of $\alpha^{(n^{}_q)}_s$ in
Eq. (9), we need to know the matching condition of the running quark
masses at the flavor thresholds \cite{Kniehl,matching}:
\begin{eqnarray}
\frac{m^{(n^{}_q-1)}_q}{m^{(n^{}_q)}_q} & = & 1 + \frac{1}{12}\left(
\frac{\alpha^{(n^{}_q)}_s(\mu)}{\pi} \right)^2\left( \ln^2
\frac{\mu^2}{\left(\mu^{(n^{}_q)} \right)^2} - \frac{5}{3}\ln
\frac{\mu^2}{\left(\mu^{(n^{}_q)}\right)^2} +\frac{89}{36}
\right) \nonumber \\
&&  + \left( \frac{\alpha^{(n^{}_q)}_s(\mu)}{\pi} \right)^3
\left[\frac{2951}{2916}
-\frac{407}{864}\zeta(3)+\frac{5}{4}\zeta(4)-\frac{1}{36}B_4 \right.
\nonumber \\
&& \left.
-\left(\frac{311}{2592}+\frac{5}{6}\zeta(3)\right)
\ln\frac{\mu^2}{\left(\mu^{(n^{}_q)}\right)^2}
+ \frac{175}{432}\ln^2 \frac{\mu^2}{\left(\mu^{(n^{}_q)}\right)^2} +
\frac{29}{216}\ln^3 \frac{\mu^2}{\left(\mu^{(n^{}_q)}\right)^2}
\right. \nonumber
\\
&& \left.
(n^{}_q-1)\left(\frac{1327}{11664}-\frac{2}{27}\zeta(3)-\frac{53}{432}
\ln\frac{\mu^2}{\left(\mu^{(n^{}_q)}\right)^2}-\frac{1}{108}\ln^3
\frac{\mu^2}{\left(\mu^{(n^{}_q)}\right)^2}\right)\right] \ ,
\end{eqnarray}
where $B_4 \simeq -1.762800$. In practical calculations, we apply
Eqs. (9) and (14) to the evolution of quark masses just at the
thresholds $\mu^{(n^{}_q)}$. Between two neighboring flavor
thresholds, the RGE given in Eq. (2) or Eq. (6) will be used. We
remark that this treatment is more reasonable than that adopted in
Ref. \cite{Koide98}.

It is worthwhile to mention that the running and decoupling of the
strong coupling and quark masses have been built into the
Mathematica package \textsl{RunDec} by Chetyrkin, K\"{u}hn and
Steinhauser \cite{Kniehl}. We use the same formulas in our
calculations and find that the results are in good agreement with
those calculated by using \textsl{RunDec}, if the decoupling scale
is chosen as $\mu^{(n^{}_q)} = m^{}_q(m^{}_q)$.

\subsection{Running Charged-lepton Masses}

We proceed to discuss the running masses of charged leptons
\cite{Hall}. The $\mu$-dependence of the fine structure constant
$\alpha$, including QCD corrections, is described by \cite{Jens}
\begin{eqnarray}
\mu^2 \frac{\partial \alpha}{\partial \mu^2} = -\frac{\alpha^2}{\pi}
\left[ \tilde{\beta}^{}_0 + \tilde{\beta}^{}_1 \left(
\frac{\alpha}{\pi}\right) + \sum^3_{i=1} \rho^{}_i
\left(\frac{\alpha^{}_s}{\pi}\right)^i\right ]\ ,
\end{eqnarray}
where
\begin{eqnarray}
~~~~~\tilde{\beta}^{}_0 & = & -\frac{1}{3}\sum_{f}Q^2_f N^f_c \ ,
\nonumber \\
\tilde{\beta}^{}_1 & = & -\frac{1}{4}\sum_{f}Q^4_f N^f_c \
\end{eqnarray}
with $Q^{}_f$ being the electric charge of a charged fermion (i.e.,
$f = e, \mu, \tau$ for charged leptons and $f= u, d, s, c, b, t$ for
quarks) and $N^f_c$ being the color factor (i.e., $N^l_c = 1$ for
charged leptons and $N^q_c = 3$ for quarks), and
\begin{eqnarray}
\rho^{}_1 &  = &  - \sum_{q}Q^2_q \ , \nonumber
\\
\rho^{}_2 & = & \sum_q Q^2_q \left( \frac{257}{46} - \frac{11}{72}
n^{}_q \right) \ , \nonumber\\
\rho^{}_3 & = &  \sum_{q}Q^2_q \left[ -\frac{10487}{1728} -
\frac{55}{18}\zeta(3) + \left( \frac{707}{864} +
\frac{55}{54}\zeta(3)\right)n^{}_q +
\frac{77}{3888} n^{2}_q \right] \nonumber \\
&& -\frac{10}{3} \left(\sum_q Q^{}_q\right)^2 \left( \frac{11}{144}
- \frac{1}{6}\zeta(3)\right)  \; .
\end{eqnarray}
The pole masses of three charged leptons can be unambiguously
measured in experiments and their values have been determined to an
unprecedented degree of precision \cite{PDG},
\begin{eqnarray}
M^{}_e & = & 0.510998918 \pm 0.000000044 ~ {\rm MeV} \; , ~~
\nonumber \\
M^{}_\mu & = & 105.6583692 \pm 0.0000094 ~ {\rm MeV} \; ,
\\
M^{}_\tau & = & 1776.99^{+0.29}_{-0.26} ~ {\rm MeV} \; . \nonumber
\end{eqnarray}
One may convert the pole mass $M^{}_l$ into the running mass $m^{}_l
(\mu)$ by using the following equation \cite{running}:
\begin{eqnarray}
m^{}_l(\mu)=M^{}_l \left\{ 1- \frac{\alpha(\mu)}{\pi} \left[
1+\frac{3}{2} \ln \left(\frac{\mu}{m^{}_l (\mu)}\right) \right]
\right\} \ ,
\end{eqnarray}
where the subscript $l$ runs over $e$, $\mu$ and $\tau$, and the
terms of ${\cal O}(\alpha^2)$ and smaller have been omitted.

\subsection{The Strategy}

The formulas given in the preceding subsections allow us to
calculate the running masses of quarks and charged leptons up to the
Fermi scale. The basic strategy of our numerical calculations is
two-fold:
\begin{enumerate}
\item We shall use the values of strong and electromagnetic coupling
constants given by the PDG at $M^{}_Z$ \cite{PDG},
\begin{eqnarray}
&& \alpha^{}_s(M^{}_Z) = 0.1176 \pm 0.002 \; , \nonumber \\
&& \alpha(M^{}_Z)^{-1} = 127.918 \pm 0.018 \; ,
\end{eqnarray}
where $M^{}_Z = 91.1876 \pm 0.0021 ~ {\rm GeV}$ is the mass of the
$Z$ boson and it lies in the range $\mu > \mu^{(5)} =
m^{}_b(m^{}_b)$. We shall only use the central value of $M^{}_Z$ in
our calculations, because its error bar is negligibly small. From
Eq. (4) with $n^{}_q = 5$ and Eq. (20), we obtain the corresponding
strong coupling constant at the bottom quark threshold
$\alpha^{(5)}_s (\mu^{(5)}) = 0.223^{+0.008}_{-0.007}$ with
$\mu^{(5)}= m^{}_b(m^{}_b)$. As for $\alpha^{-1}$, we can directly
compute its values at some typical energy scales by using Eq. (15).
Our numerical results, which will be used in the subsequent
calculations, are listed in TABLE I.

\item By using the RGEs with $n^{}_q = 4$, we are able to run three
light quark masses and three charged-lepton masses to the first
heavy flavor threshold $\mu^{(5)} $. Then we implement the
matching conditions of $\alpha^{}_s$ and $m^{}_q$ to cross this
threshold. There is no flavor threshold on the way from
$\mu^{(5)}$ to $M^{}_Z$, and thus the relevant equations with
$n^{}_q = 5$ can be used for numerical calculations. We have also
calculated the running masses $m^{}_Q(\mu)$ of heavy quarks $Q$ at
the scale $\mu \ll \mu^{(n^{}_Q)} = m^{}_Q(m^{}_Q)$, which is
below the flavor threshold $\mu^{(n^{}_Q)}$. In our evaluation,
the relevant equations with $n^{}_Q$ active quark flavors have
been used. For example, the running top quark mass at the Fermi
scale $m^{}_t(M^{}_Z)$ is actually computed by using the RGEs with
$n^{}_q = 6$.
\end{enumerate}
Our numerical results for the running masses of quarks and charged
leptons below the Fermi scale are given in TABLE II and TABLE III,
respectively. We do not consider the running masses of three light
neutrinos in this energy region, just because their interactions
with other particles are too weak and their pole masses are too
tiny. On the one hand, one expects that the changes of neutrino
masses with respect to the energy scales are negligibly small. On
the other hand, it is just the smallness of neutrino masses that
hints at the possible existence of certain new physics scales
(e.g., the seesaw scales $\mu \sim 10^9$ GeV to $10^{12}~{\rm
GeV}$). We shall illustrate the running effects of three neutrino
masses from $M^{}_Z$ to $\mu \sim 10^{12}$ GeV by using the RGEs
of both the SM and the MSSM in the next section.

\section{Above the Fermi Scale}

Above the Fermi scale, the unbroken electroweak gauge symmetry
forbids leptons and quarks to acquire their masses. The actual
meaning of a fermion mass $m^{}_f$ in this energy region is a
measure of the nontrivial Yukawa coupling eigenvalue $y^{}_f$. One
commonly defines $m^{}_f = y^{}_f v$ above the Fermi scale, just
like the definition below the Fermi scale, where $v \approx 246$
GeV is the vacuum expectation value of the neutral Higgs field in
the SM. In the conventional seesaw models \cite{SS}
\footnote{Here we follow the widely-accepted and well-motivated
assumption that massive neutrinos are Majorana particles. The RGE
running effects of Dirac neutrino masses and flavor mixing
parameters have been studied in Ref. \cite{Dirac}.},
the heavy Majorana neutrinos must be integrated out below the seesaw
scale and the effective coupling matrix of three light Majorana
neutrinos is given by the well-known seesaw relation $\kappa =
-Y^{}_\nu M^{-1}_{\rm R} Y^{T}_\nu$, where $Y^{}_\nu$ denotes the
neutrino Yukawa coupling matrix and $M^{}_{\rm R}$ is the
right-handed Majorana neutrino mass matrix. The light neutrino
masses are therefore given by $m^{}_i = \kappa^{}_i v^2$ with
$\kappa^{}_i$ (for $i=1, 2, 3$) being the eigenvalues of $\kappa$.
Above the seesaw scale, the flavor threshold effects induced by the
masses of heavy Majorana neutrinos have to be carefully treated and
the details of $Y^{}_\nu$ have to be model-dependently assumed
\cite{Mei}. Hence we shall only evaluate the running neutrino masses
$m^{}_i$ from the Fermi scale up to the seesaw scale, in order to
avoid the complications and uncertainties associated with the seesaw
thresholds.

Then what we are concerned with is the evolution of four Yukawa
coupling matrices $Y^{}_u$, $Y^{}_d$, $Y^{}_l$ and $\kappa$, whose
eigenvalues correspond to the masses of up-type quarks ($m^{}_u,
m^{}_c, m^{}_t$), down-type quarks ($m^{}_d, m^{}_s, m^{}_b$),
charged leptons ($m^{}_e, m^{}_\mu, m^{}_\tau$) and neutrinos
($m^{}_1, m^{}_2, m^{}_3$). Their one-loop RGEs in the SM and MSSM
can be found in Refs. \cite{Dirac,Mei}. The two-loop beta functions
of $Y^{}_u$, $Y^{}_d$ and $Y^{}_l$ have already been derived in Ref.
\cite{2loopRGE}. Their lengthy expressions will not be quoted here,
but they will be used in our numerical calculations. As for
neutrinos, we only consider the one-loop RGEs for their running
masses because our present knowledge on the absolute values of
neutrino masses remain quite limited and uncertain \cite{PDG}. The
strategy of computing the running quark and lepton masses above the
Fermi scale is outlined as follows. First, we use the quark masses
and flavor mixing parameters obtained at the Fermi scale to
reconstruct the Yukawa coupling matrices $Y^{}_u$ and $Y^{}_d$.
Second, the RGEs of $Y^{}_u$ and $Y^{}_d$ are solved and their
eigenvalues are determined to give the running quark masses at every
energy scale of our interest. We may follow a similar procedure to
reconstruct $Y^{}_l$ and $\kappa$ from current experimental data at
$M^{}_Z$, and then we solve their RGEs to find out the running
lepton masses above $M^{}_Z$. Because the RGEs of $Y^{}_u$,
$Y^{}_d$, $Y^{}_l$ and $\kappa$ are more or less entangled, our
numerical calculations will be done simultaneously for quarks and
leptons.

Now that the running quark masses are to be evaluated in the
supersymmetric framework, the matching procedure from the SM to the
MSSM has to be taken into account. The RGEs in a supersymmetric
theory are usually derived in the $\overline{\rm DR}$ scheme based
on the dimensional reduction and the minimal subtraction \cite{DR},
while the experimental data on quark masses are extracted by using
the $\overline{\rm MS}$ scheme. Hence the transition between these
two schemes needs to be treated in a consistent way
\cite{MS2DR,Steinhauser}. To get around the occurrence of
intermediate non-supersymmetric effective theories, here we follow
Ref. \cite{Steinhauser} to adopt the {\it common scale approach}
with all the supersymmetric particles being roughly at a common
scale $\tilde{M}$. In our analysis, we set the decoupling scale to
be $\mu=\tilde{M}=M_Z$. Furthermore, the input values of
$\alpha^{}_s(M^{}_Z)$ and $m^{}_q(M^{}_Z)$ in the $\overline{\rm
MS}$ scheme will be converted into those in the $\overline{\rm DR}$
scheme at $M^{}_Z$. First, since the two-loop RGEs will be used
above the Fermi scale, we consider the transition of $\alpha^{}_s$
from the $\overline{\rm MS}$ scheme to the $\overline{\rm DR}$
scheme at the one-loop level \cite{alphasDR3},
\begin{eqnarray}
\alpha^{\overline{\rm MS}}_s = \alpha^{\overline{\rm DR}}_s
\left(1-\frac{\alpha^{\overline{\rm DR}}_s}{4\pi}\right) \ .
\end{eqnarray}
Given the input value of $\alpha^{}_s(M_Z)$ in Eq. (20),
$\alpha^{\overline{\rm DR}}_s(M_Z) \simeq 0.1187$ can be obtained
and will be used as the input value in our numerical calculations.
We take account of the matching between the SM and MSSM at $M^{}_Z$;
i.e., $\alpha^{{\rm SM}}_s = \zeta^{}_s \alpha^{\rm MSSM}_s$. In the
common scale approach, the decoupling constant approximates to
$\zeta^{}_s \approx 1$ at the one-loop level, implying that the top
quark and heavy sparticles are simultaneously decoupled. Second, the
one-loop matching of quark masses between the $\overline{\rm MS}$
and $\overline{\rm DR}$ schemes is simply given by
\begin{eqnarray}
m^{\overline{\rm DR}}_q=m^{\overline{\rm MS}}_q \left(1-
\frac{\alpha^{}_e}{3\pi}\right) \ ,
\end{eqnarray}
where $\alpha^{}_e$ is the evanescent coupling appearing in a
non-supersymmetric theory renormalized in the $\overline{\rm DR}$
scheme. We have $\alpha^{}_e \approx \alpha^{\overline{\rm DR}}_s$
up to ${\cal O}(\alpha^2_s)$ \cite{Steinhauser}. It is well known
that the matching effects between the SM and MSSM are only
significant for the masses of down-type quarks and charged leptons
because of the large-$\tan \beta$ enhancement
\cite{Baer,Serna,Antusch,Buras}. At the one-loop level, the
threshold corrections read
\begin{eqnarray}
m^{\rm MSSM}_i=\frac{m^{\rm SM}_i}{1+\epsilon^{}_i \tan\beta} \ ,
\end{eqnarray}
where $i=d,s,b$ for down-type quarks or $i=e,\mu,\tau$ for charged
leptons, and the definitions of $\epsilon^{}_i$ can be found in
Refs. \cite{Antusch,Buras}. In our numerical analysis with the
common scale approach, $\epsilon^{}_i$ can be as large as one
percent. Hence the supersymmetric threshold corrections are
particularly relevant in the case of sizable $\tan\beta$ (e.g.,
$\tan \beta = 50$). It is worth remarking that we have tried to
avoid the details of any specific supersymmetric models within the
scope of this work. If they are taken into account, however, a more
careful treatment of the decoupling of supersymmetric partners will
be unavoidable \cite{Baer,Serna,Antusch}.

Before doing the numerical calculations, let us briefly summarize
the relevant data to be input at the Fermi scale.
\begin{enumerate}
\item In the basis where the Yukawa coupling matrix $Y^{}_u$ is
diagonal (i.e., $Y^{}_u = {\rm Diag}\{y^{}_u, y^{}_c, y^{}_t\}$
with $y^{}_q = m^{}_q/v$ for $q=u, c, t$), the mass eigenstates of
down-type quarks $(d, s, b)$ are related to their weak eigenstates
$(d', s', b')$ by the unitary Cabibbo-Kobayashi-Maskawa (CKM)
matrix $V$ \cite{CKM}. Namely, $V^\dagger Y^{}_dY^\dagger_d V =
{\rm Diag} \{y^2_d, y^2_s, y^2_b \}$ with $y^{}_q = m^{}_q/v$ for
$q=d, s$ and $b$. It is therefore possible to reconstruct $Y^{}_d$
from its eigenvalues and $V$. Because quarks are Dirac particles,
the CKM matrix $V$ can be parametrized in terms of four
independent quantities, such as the moduli of three independent
elements of $V$ and one CP-violating phase \cite{FX95}. For
example, one may choose the following set of parameters, which
have been precisely measured, to parametrize $V$ at $M^{}_Z$
\cite{PDG}:
\begin{eqnarray}
&& \left| V^{}_{cb} \right| = (41.6 \pm 0.6) \times 10^{-3} \; ,
\nonumber \\
&& \left| V^{}_{us} \right| = 0.2257 \pm 0.0021 \; , \nonumber \\
&& \left| V^{}_{ub} \right| = (4.31\pm0.30)\times 10^{-3} \; ,
\nonumber \\
&& \sin 2\beta  =  0.687 \pm 0.032 \;,
\end{eqnarray}
where $\beta \equiv \arg
\left[-(V^{}_{cd}V^{\ast}_{cb})/(V^{}_{td}V^{\ast}_{tb}) \right]$ is
an inner angle of the CKM unitarity triangle. We shall only use the
central values of the above parameters in our numerical
calculations, because the running quark and lepton masses are
actually insensitive to these inputs. On the other hand, the values
of quark masses at the Fermi scale have been evaluated in Sec. II
and listed in TABLE II.

\item Without loss of generality, we choose the flavor basis
where the Yukawa coupling matrix $Y^{}_l$ is diagonal (i.e.,
$Y^{}_l = {\rm Diag} \{y^{}_e, y^{}_\mu, y^{}_\tau \}$ with
$y^{}_l = m^{}_l/v$ for $l= e, \mu, \tau$). In this basis, the
mass eigenstates of three light Majorana neutrinos ($\nu^{}_1,
\nu^{}_2, \nu^{}_3$) are linked to their weak eigenstates
($\nu^{}_e, \nu^{}_\mu, \nu^{}_\tau$) through the unitary
Maki-Nakagawa-Sakata (MNS) matrix $U$ \cite{MNS}
\footnote{Because the seesaw scales are assumed to be extremely
higher than the Fermi scale in this paper, $U$ is expected to be
unitary as an excellent approximation. See, e.g., Ref. \cite{NU}
for a detailed analysis of the non-unitary neutrino mixing in the
seesaw scenarios.}.
Namely, $U^\dagger \kappa U^* = {\rm Diag}\{ \kappa^{}_1,
\kappa^{}_2, \kappa^{}_3 \}$ with $\kappa^{}_i = m^{}_i/v^2$ for
$i=1, 2$ and $3$. It is then easy to reconstruct the symmetric
matrix $\kappa$ from $m^{}_i$ and $U$ at the Fermi scale. A
complete parametrization of the MNS matrix $U$ needs three mixing
angles and three CP-violating phases \cite{Xing04},
\begin{eqnarray}
U = \left( \matrix{c^{}_{12} c^{}_{13} & s^{}_{12} c^{}_{13} &
s^{}_{13} e^{-i\delta^{}_l} \cr -s^{}_{12} c^{}_{23}-c^{}_{12}
s^{}_{23} s^{}_{13}
 e^{i \delta^{}_l} & c^{}_{12} c^{}_{23}-s^{}_{12} s^{}_{23} s^{}_{13}
 e^{i \delta^{}_l} & s^{}_{23} c^{}_{13} \cr
 s^{}_{12} s^{}_{23}-c^{}_{12} c^{}_{23} s^{}_{13}
 e^{i \delta^{}_l} & -c^{}_{12} s^{}_{23}-s^{}_{12} c^{}_{23} s^{}_{13}
 e^{i \delta^{}_l} & c^{}_{23} c^{}_{13}}
\right)\left( \matrix{ e^{i \rho} & 0 & 0 \cr 0 & e^{i \sigma} & 0
\cr 0 & 0 & 1 } \right)
\end{eqnarray}
with $c^{}_{ij} \equiv \cos \theta^{}_{ij}$ and $s^{}_{ij} \equiv
\sin \theta^{}_{ij}$ (for $ij=12$, $13$ and $23$). A global analysis
of current neutrino oscillation data yields the constraints
\cite{SV}
\begin{eqnarray}
30^{\circ} & < & \theta^{}_{12} ~ < ~ 38^{\circ} \ , \nonumber \\
36^{\circ} & < & \theta^{}_{23} ~ < ~ 54^{\circ} \ , \nonumber \\
\theta^{}_{13} & < & 10^{\circ}
\end{eqnarray}
at the $99\%$ confidence level. Three CP-violating phases
$\delta^{}_{l}$, $\rho$ and $\sigma$ remain entirely unrestricted.
It has been noticed that the running behaviors of neutrino masses
are essentially insensitive to three neutrino mixing angles and
three CP-violating phases in the SM or in the MSSM with small
$\tan\beta$ \cite{Mei}. Therefore, we shall simply assume
$\delta^{}_l = \rho = \sigma =0$ and take the best-fit values
$\theta^{}_{12} \approx 33.8^{\circ}$, $\theta^{}_{23} \approx
45^{\circ}$ and $\theta^{}_{13} \approx 0^{\circ}$ as the typical
inputs at $M^{}_{Z}$. Note that the absolute values of $m^{}_i$
remain unknown, although their upper bound is expected to be ${\cal
O}(1)$ eV. For illustration, we only consider two possibilities in
our numerical calculations: (A) the normal neutrino mass hierarchy
with $m^{}_1 = 0.001 ~{\rm eV}$ and $m^{}_1 < m^{}_2 \ll m^{}_3$;
and (B) the nearly degenerate neutrino mass spectrum with $m^{}_1 =
0.2 ~ {\rm eV}$ and $m^{}_1 \lesssim m^{}_2 \lesssim m^{}_3$. The
best-fit values $\Delta m^2_{21} = 8.0 \times 10^{-5} ~ {\rm eV^2}$
and $\Delta m^2_{32} = 2.5 \times 10^{-3} ~ {\rm eV^2}$ are input at
$M^{}_Z$ in both cases. This simplified treatment can easily be
improved in the future, once more experimental data on $m^{}_i$ and
$U$ are available.

\item In the SM or MSSM with
the gauge group $SU(3)^{}_{\rm C} \times SU(2)^{}_{\rm L} \times
U(1)^{}_{\rm Y}$, three gauge couplings $g^{}_s$, $g$ and $g'$ are
given by
\begin{eqnarray}
&& g^{2}_s = 4 \pi \alpha^{}_{s} \; , \nonumber \\
&& g^2 = 4 \pi \alpha/\sin^2\theta^{}_{W} \; , \nonumber \\
&& g' = g \tan \theta^{}_{W} \ ,
\end{eqnarray}
where $\theta^{}_{W}$ is the weak mixing angle. In the $\rm SU(5)$
GUTs with or without supersymmetry, the gauge coupling constants
$g^{}_1$, $g^{}_2$ and $g^{}_3$ are usually normalized as $g^{}_3 =
g^{}_s$, $g^{}_2 = g$ and $g^{}_{1} = \sqrt{5/3} ~ g'$. The RGEs of
$g^{}_i$ (for $i=1, 2, 3$) are given in Ref. \cite{2loopRGE}. The
input parameters include \cite{PDG}
\begin{eqnarray}
\sin^2\theta^{}_{W}(M^{}_Z) & = & 0.23122 \pm 0.00015 \ ,
\end{eqnarray}
as well as $\alpha^{}_{s}(M^{}_Z)$ and $\alpha(M^{}_{Z})$ given in
Eq. (20). Besides the gauge couplings, the quartic Higgs coupling
$\lambda = 2m^2_H/v^2$ in the SM is also needed in our calculations.
The RGE of $\lambda$ can be found in Ref. \cite{2loopRGE}. We shall
typically take $m^{}_{H} = 140 ~{\rm GeV}$ for the Higgs mass, just
for the sake of illustration. In the MSSM, the free parameter
$\tan\beta$ is defined as $\tan \beta = v^{}_u/v^{}_d$ with
$v^{}_{d}$ and $v^{}_u$ being the vacuum expectation values of the
Higgs fields which couple respectively to the down- and up-type
quarks. Note that $v^2 = v^2_d + v^2_u = 4M^2_{W}/g^2 \approx
(246~{\rm GeV})^2$ has been fixed by the measurements of the
$W$-boson mass and the electroweak gauge coupling. To be specific,
we shall input $\tan\beta = 10$ and $\tan\beta = 50$ to illustrate
the running fermion masses in the cases of small and large $\tan
\beta$.
\end{enumerate}
With the help of the above inputs, we are then able to numerically
solve the RGEs and obtain the running masses of quarks and leptons
at various energies above the Fermi scale.

Our numerical results at $\mu = 1 ~ {\rm TeV}$, $10^9 ~ {\rm GeV}$,
$10^{12} ~ {\rm GeV}$ and $\Lambda^{}_{\rm GUT} \sim 2 \times
10^{16}~ {\rm GeV}$ are summarized in TABLE IV, TABLE V and TABLE
VI, where the uncertainties of the output masses mainly come from
those of the input masses at $M^{}_Z$. Some brief comments on these
results are in order.
\begin{itemize}
\item In both the SM and the MSSM, a common feature of the outputs
is that all the quark masses decrease with increasing energy scales.
Nevertheless, their running behaviors are quite model-dependent.

\item TABLE IV and TABLE VI show that there exists a
maximum for the running mass of each charged lepton in the SM. Our
numerical analysis indicates that the maximal values of $m^{}_e$,
$m^{}_\mu$ and $m^{}_\tau$ are all located around $\mu \sim 10^6 ~
{\rm GeV}$. In the MSSM, however, three charged-lepton masses
smoothly decrease as the energy scale increases.

\item We have ignored large uncertainties of the input values in evaluating
the running masses of three light neutrinos, just because the
relevant experimental data are quite inaccurate and incomplete.
Our numerical results, which rely on several assumptions made
above, can only serve for illustration. We show the scale
dependence of running neutrino masses in FIGs. 1 and 2 for two
different mass spectra. Comparing between these two figures, we
observe that the RGE effects on neutrino masses are more
significant in the $m^{}_1 \lesssim m^{}_2 \lesssim m^{}_3$ case
than in the $m^{}_1 < m^{}_2 \ll m^{}_3$ case. Similar
observations have been made by a number of authors before (see,
e.g., Ref. \cite{Mei}).
\end{itemize}
Note that the running behaviors of other parameters, such as the CKM
and MNS mixing angles, CP-violating phases and gauge coupling
constants, can be obtained from our program in a straightforward
way. But here we only concentrate on the outputs of running quark
and lepton masses.

\section{summary}

In this paper, we have updated the running masses of quarks and
leptons both in the SM and in the MSSM at various energy scales,
including $\mu = M^{}_Z$, $1~{\rm TeV}$, $10^9 ~ {\rm GeV}$,
$10^{12} ~ {\rm GeV}$ and $\Lambda^{}_{\rm GUT}$. Our motivation
is simple but meaningful: we want to provide a reliable and
up-to-date table of the running quark and lepton masses for
particle physicists. Such a table will be very useful for the
analysis of hadronic physics at low energies and for the building
of new physics models at superhigh energies.

The differences between our work and the previous works (in
particular, Ref. \cite{Koide98}) have been summarized in the
introductory section of this paper. For instance, the central
value of the strange quark mass $m^{}_s$ at $M^{}_Z$ is about 55
MeV today, but it was about 93 MeV as given in Ref. \cite{Koide98}
about a decade ago. Hence it makes sense to recalculate the values
of $m^{}_s$ and other fermion masses at different energy scales.
In particular, we hope that our new results for the running masses
of leptons and quarks may help the model builders to get new
insight into the flavor dynamics at the energy frontier set by the
LHC and in the exciting era of precision neutrino physics.

\acknowledgments{This work is supported in part by the National
Natural Science Foundation of China.}

\newpage

\begin{table}
\caption{The gauge couplings $\alpha^{}_s(\mu)$ and
$\alpha(\mu)^{-1}$ at a few typical energy scales, where the first
and second errors of $\alpha(\mu)^{-1}$ come from the uncertainties
associated with $\alpha(M^{}_Z)^{-1}$ and $m^{}_q(m^{}_q)$,
respectively.}
\begin{center}
\begin{tabular}{c|c|c|c}
$ n^{}_q $ & $\mu^{(n^{}_q)}=m^{}_q(m^{}_q) ~ ({\rm GeV})$ & 
$\alpha^{}_s(\mu)$ & $\alpha(\mu)^{-1}$ \\
\hline
&&&\\
4 & $m^{}_c(m^{}_c) = 1.25 $ & 
$0.387^{+0.027}_{-0.024}$ & $134.116 \pm 0.018^{+0.109}_{-0.101}$ \\
\hline
&&&\\
5 & $m^{}_b(m^{}_b) = 4.20$ & 
$0.223^{+0.008}_{-0.007}$ & $132.406 \pm 0.018^{+0.024}_{-0.025}$ \\
\hline
&&&\\
6 & $m^{}_t(m^{}_t) = 163.6 $ & 
$0.108 \pm 0.002$ & $127.073 \pm 0.018 \pm {0.023}$ \\
\end{tabular}
\end{center}
\end{table}

\begin{table}
\caption{Running quark masses from $\mu^{(4)}=m^{}_c(m^{}_c)$ to
$\mu^{(6)}=m^{}_t(m^{}_t)$, where we have taken the $W$-boson mass
to be $M^{}_W = 80.403~{\rm GeV}$. The pole masses of three light
quarks are not listed, simply because the perturbative QCD
calculation is not reliable in that energy region. }
\begin{center}
\begin{tabular}{c|c|c|c|c|c|c}
$\mu$ & $m^{}_u(\mu)~ ({\rm MeV})$ &  $m^{}_d(\mu)~ ({\rm MeV})$ & $m^{}_s(\mu)~ ({\rm MeV})$ & $m^{}_c(\mu)~ ({\rm GeV})$ & $m^{}_b(\mu)~ ({\rm GeV})$ & $m^{}_t(\mu)~ ({\rm GeV})$ \\
\hline
&&&&&\\
$m^{}_c(m^{}_c)$ & $2.57^{+0.99}_{-0.84}$ & $5.85^{+2.46}_{-2.38}$ & $111^{+31}_{-30}$ & $1.25 \pm 0.09$ & $5.99^{+0.28}_{-0.26}$ & $384.8^{+22.8}_{-20.4}$\\
\hline
&&&&&\\
$2 ~{\rm GeV}$ & $2.2^{+0.8}_{-0.7}$ & $5.0 \pm 2.0$  & $95 \pm 25$ & $1.07^{+0.12}_{-0.13}$ & $5.05^{+0.16}_{-0.15}$ & $318.4^{+13.3}_{-12.4}$ \\
\hline
&&&&&\\
$m^{}_b(m^{}_b)$ & $1.86^{+0.70}_{-0.60}$ & $4.22^{+1.74}_{-1.71}$ & $80 \pm 22$ & $0.901^{+0.111}_{-0.113}$ & $4.20 \pm 0.07$ & $259.8^{+7.7}_{-7.4}$ \\
\hline
&&&&&\\
$M^{}_W$ & $1.29^{+0.50}_{-0.43}$ & $2.93^{+1.25}_{-1.21}$ & $56 \pm 16$ & $0.626^{+0.084}_{-0.085} $ & $2.92 \pm 0.09$ & $173.8 \pm 3.0$  \\
\hline
&&&&&\\
$M^{}_Z$ & $1.27^{+0.50}_{-0.42}$ & $2.90^{+1.24}_{-1.19}$ & $55^{+16}_{-15}$ & $0.619 \pm 0.084$ & $2.89 \pm 0.09$ & $171.7 \pm 3.0$ \\
\hline
&&&&&\\
$m^{}_t(m^{}_t)$ & $1.22^{+0.48}_{-0.40}$ & $2.76^{+1.19}_{-1.14}$ & $52 \pm 15$ & $0.590 \pm 0.080$ & $2.75 \pm 0.09$ & $162.9 \pm 2.8$ \\
\hline
&&&&&\\
$M^{}_q$ & $\sim$ & $\sim$ & $\sim$ & $1.77 \pm 0.14$ & $4.91^{+0.12}_{-0.11}$ & $172.5 \pm 2.7$ \\
\hline
&&&&&\\
$m^{}_q(M^{}_q)$ & $\sim$ & $\sim$ & $\sim$ & $1.11^{+0.11}_{-0.12}$ & $4.08 \pm 0.08$ & $162.2 \pm 2.8 $ \\
\end{tabular}
\end{center}
\end{table}

\begin{table}
\caption{Running charged-lepton masses below $m^{}_t(m^{}_t)$, where
the uncertainties of $m^{}_l(\mu)$ are determined by those of
$M^{}_l$.}
\begin{center}
\begin{tabular}{c|c|c|c}
$\mu$ & $m^{}_e(\mu)~ ({\rm MeV})$ &  $m^{}_\mu(\mu)~ ({\rm MeV})$ & $m^{}_\tau(\mu)~ ({\rm MeV})$ \\
\hline
&&&\\
$m^{}_c(m^{}_c)$ & $0.495536319 \pm {+0.000000043}$ & $104.4740056^{+0.0000093}_{-0.0000094}$ & $1774.90 \pm 0.20$ \\
\hline
&&&\\
$m^{}_b(m^{}_b)$ & $0.493094195^{+0.000000042}_{-0.000000043}$ & $103.9951891 \pm {0.0000093}$ & $1767.08 \pm 0.20$ \\
\hline
&&&\\
$M^{}_W$ & $0.486845675 \pm 0.000000042 $ & $102.7720886 \pm 0.0000092$ & $1747.12^{+0.20}_{-0.19}$ \\
\hline
&&&\\
$M^{}_Z$ & $0.486570161 \pm 0.000000042 $ & $102.7181359 \pm 0.0000092 $ & $1746.24^{+0.20}_{-0.19}$ \\
\hline
&&&\\
$m^{}_t(m^{}_t)$ & $0.485289396 \pm 0.000000042 $ & $102.4673155^{+0.0000091}_{-0.0000092}$  & $1742.15 \pm 0.20$ \\
\hline
&&&\\
$M^{}_l$ & $0.510998918 \pm 0.000000044 $  & $105.6583692 \pm 0.0000094$ & $1776.90 \pm 0.20$ \\
\end{tabular}
\end{center}
\end{table}

\begin{table}
\caption{Running quark and lepton masses above $M^{}_Z$ in the SM
with $m^{}_H = 140$ GeV, where the uncertainties of $m^{}_f(\mu)$
result from those of $m^{}_f(M^{}_Z)$. Here we have used
$\Lambda^{}_{\rm GUT} = 2\times10^{16} ~ {\rm GeV}$. Case A and case
B represent two different neutrino mass patterns with
$m^{}_1(M^{}_Z) = 0.001~{\rm eV}$ and $m^{}_1(M^{}_Z) = 0.2~{\rm
eV}$, respectively.}
\begin{center}
\begin{tabular}{l|l|l|l|l|l}
& $\mu=M^{}_Z $ & $\mu=1~ {\rm TeV}$ & $\mu=10^9~{\rm GeV}$ & $\mu=10^{12}~{\rm GeV}$ & $\mu=\Lambda^{}_{\rm GUT}$  \\
\hline
&&&&&\\
$m^{}_u(\mu)~(\rm MeV)$ &  $1.27^{+0.50}_{-0.42}$ & $1.10^{+0.43}_{-0.37}$ &  $0.67^{+0.27}_{-0.23}$ & $0.58^{+0.24}_{-0.20}$ & $0.48^{+0.20}_{-0.17}$ \\
\hline
&&&&&\\
$m^{}_d(\mu)~(\rm MeV)$ & $2.90^{+1.24}_{-1.19}$ & $2.50^{+1.08}_{-1.03}$ & $1.56^{+0.69}_{-0.65}$ & $1.34^{+0.60}_{-0.56}$ & $1.14^{+0.51}_{-0.48}$\\
\hline
&&&&&\\
$m^{}_s(\mu)~(\rm MeV)$ & $55^{+16}_{-15}$ & $47^{+14}_{-13}$ & $30^{+9}_{-8}$ & $26^{+8}_{-7}$ & $22^{+7}_{-6}$ \\
\hline
&&&&&\\
$m^{}_c(\mu)~(\rm GeV)$ & $0.619 \pm 0.084$ & $0.532^{+0.074}_{-0.073} $ & $0.327^{+0.048}_{-0.047}$ & $0.281^{+0.042}_{-0.041}$ & $0.235^{+0.035}_{-0.034}$\\
\hline
&&&&&\\
$m^{}_b(\mu)~(\rm GeV)$ &  $2.89 \pm 0.09$ & $2.43\pm 0.08$ & $1.42 \pm 0.06$ & $1.21 \pm 0.05$ & $1.00 \pm 0.04$\\
\hline
&&&&&\\
$m^{}_t(\mu)~(\rm GeV)$ & $171.7 \pm 3.0$  & $150.7 \pm 3.4 $ & $99.1^{+4.0}_{-3.8} $ & $86.7^{+4.0}_{-3.8}$ & $74.0^{+4.0}_{-3.7}$\\
\hline
 & $0.486570161$ & $0.495901601$ & $0.501014122$ & $0.490856087$ & $0.469652046$ \\
$m^{}_e(\mu)~(\rm MeV)$ & $\pm {0.000000042}$ & $\pm {0.000000043}$ & $\pm {0.000000043}$ & ${}^{+0.000000042}_{-0.000000043}$ & $\pm {0.000000041}$ \\
\hline
& $102.7181359 $& $104.6880645$ & $105.7673562$ & $103.6229311$ & $99.1466226$\\
$m^{}_\mu(\mu)~(\rm MeV)$ & $\pm 0.0000092 $ & ${}^{+0.0000094}_{-0.0000093}$ & ${}^{+0.0000095}_{-0.0000094}$  & ${}^{+0.0000092}_{-0.0000093}$ & $\pm 0.0000089 $ \\
\hline
&&&&&\\
$m^{}_\tau(\mu)~(\rm MeV)$ &$1746.24^{+0.20}_{-0.19}$ & $1779.74 \pm 0.20$ & $1798.11^{+0.21}_{-0.20}$ & $1761.67 \pm 0.20$ & $1685.58 \pm 0.19$\\
\hline
\\
{\bf Case A}  \\
\hline
&&&&&\\
$m^{}_1(\mu)~({\rm eV})$ & $0.001$ & $0.001$  & $0.001$  & $0.001$ & $\sim$ \\
\hline
&&&&&\\
$\Delta m^{2}_{21}(\mu)~({\rm eV^2})$  & $8.0\times 10^{-5}$ & $9.1\times 10^{-5}$  & $1.3\times 10^{-4}$ & $1.5\times 10^{-4}$ & $\sim$ \\
\hline
&&&&&\\
$\Delta m^{2}_{32}(\mu)~({\rm eV^2})$  & $2.5\times 10^{-3}$ & $2.9\times 10^{-3}$ & $4.2\times 10^{-3}$ & $4.6\times 10^{-3}$ & $\sim$ \\
\hline
\\
{\bf Case B} \\
\hline
&&&&&\\
$m^{}_1(\mu)~({\rm eV})$ & $0.20$ & $0.21$  & $0.26$  & $0.27$ & $\sim$ \\
\hline
&&&&&\\
$\Delta m^{2}_{21}(\mu)~({\rm eV^2})$  & $8.0\times 10^{-5}$ & $9.1\times 10^{-5}$  & $1.3\times 10^{-4}$ & $1.5\times 10^{-4}$ & $\sim$ \\
\hline
&&&&&\\
$\Delta m^{2}_{32}(\mu)~({\rm eV^2})$  & $2.5\times 10^{-3}$ & $2.9\times 10^{-3}$ & $4.2\times 10^{-3}$ & $4.6\times 10^{-3}$ & $\sim$ \\
\end{tabular}
\end{center}
\end{table}

\begin{table}
\caption{Running quark and lepton masses above $M^{}_Z$ in the MSSM
with $\tan\beta=10$, where the matching effect between the MS and
MSSM and the $\overline{\rm MS}$-to-$\overline{\rm DR}$ transition
effect on the input parameters at $M^{}_Z$ have been taken into
account.}
\begin{center}
\begin{tabular}{l|l|l|l|l|l}
& $\mu=M^{}_Z $ & $\mu=1~ {\rm TeV}$ & $\mu=10^9~{\rm GeV}$ & $\mu=10^{12} ~{\rm GeV}$ & $\mu=\Lambda^{}_{\rm GUT}$  \\
\hline
&&&&&\\
$m^{}_u(\mu)~(\rm MeV)$ &  $1.27^{+0.50}_{-0.42}$ & $1.15^{+0.45}_{-0.38}$ &  $0.75^{+0.30}_{-0.25}$ & $0.62^{+0.26}_{-0.21}$ & $0.49^{+0.20}_{-0.17}$ \\
\hline
&&&&&\\
$m^{}_d(\mu)~(\rm MeV)$ & $2.90^{+1.24}_{-1.19}$ & $2.20^{+0.96}_{-0.91}$ & $1.21^{+0.54}_{-0.51}$ & $0.96^{+0.43}_{-0.40}$ & $0.70^{+0.31}_{-0.30}$\\
\hline
&&&&&\\
$m^{}_s(\mu)~(\rm MeV)$ & $55^{+16}_{-15}$ & $42 \pm 12$ & $23 \pm 7$ & $18^{+6}_{-5}$ & $13 \pm {4}$ \\
\hline
&&&&&\\
$m^{}_c(\mu)~(\rm GeV)$ & $0.619 \pm 0.084$ & $0.557^{+0.077}_{-0.076} $ & $0.363^{+0.053}_{-0.052}$ & $0.303^{+0.046}_{-0.045}$ & $0.236^{+0.037}_{-0.036}$\\
\hline
&&&&&\\
$m^{}_b(\mu)~(\rm GeV)$ &  $2.89 \pm 0.09$ & $2.23\pm 0.08$ & $1.30 \pm 0.05$ & $1.05 \pm 0.05$ & $0.79 \pm 0.04$\\
\hline
&&&&&\\
$m^{}_t(\mu)~(\rm GeV)$ & $171.7 \pm 3.0$  & $161.0^{+3.7}_{-3.6}$ & $125.2^{+7.1}_{-6.5} $ & $111.0^{+8.5}_{-7.4}$ & $92.2^{+9.6}_{-7.8}$\\
\hline
 & $0.486570161$ & $0.418436115$ & $0.358332424$ & $0.327996884$ & $0.283755495$ \\
$m^{}_e(\mu)~(\rm MeV)$ & $\pm {0.000000042}$ & $\pm {0.000000036}$ & $\pm {0.000000031}$ & ${}^{+0.000000028}_{-0.000000029}$ & ${}^{+0.000000024}_{-0.000000025}$ \\
\hline
& $102.7181359 $& $88.3347018$ & $75.6468538$ & $69.2429377$ & $59.9033617$\\
$m^{}_\mu(\mu)~(\rm MeV)$ & $\pm 0.0000092 $ & $\pm {0.0000079}$ & $\pm 0.0000068 $ & $\pm 0.0000062 $ & $\pm 0.0000054 $  \\
\hline
&&&&&\\
$m^{}_\tau(\mu)~(\rm MeV)$ &$1746.24^{+0.20}_{-0.19}$ & $1502.25\pm{0.17}$ & $ 1288.68 \pm {0.15}$ & $1180.38^{+0.13}_{-0.14} $ & $1021.95^{+0.11}_{-0.12}$\\
\hline
\\
{\bf Case A}  \\
\hline
&&&&&\\
$m^{}_1(\mu)~({\rm eV})$ & $0.001$ & $0.001$  & $0.001$  & $0.001$ & $\sim$ \\
\hline
&&&&&\\
$\Delta m^{2}_{21}(\mu)~({\rm eV^2})$  & $8.0\times 10^{-5}$ & $8.7\times 10^{-5}$  & $1.0\times 10^{-4}$ & $1.0\times 10^{-4}$ & $\sim$ \\
\hline
&&&&&\\
$\Delta m^{2}_{32}(\mu)~({\rm eV^2})$  & $2.5\times 10^{-3}$ & $2.7\times 10^{-3}$ & $3.3\times 10^{-3}$ & $3.1\times 10^{-3}$ & $\sim$ \\
\hline
\\
{\bf Case B} \\
\hline
&&&&&\\
$m^{}_1(\mu)~({\rm eV})$ & $0.20$ & $0.21$  & $0.23$  & $0.22$ & $\sim$ \\
\hline
&&&&&\\
$\Delta m^{2}_{21}(\mu)~({\rm eV^2})$ & $8.0\times 10^{-5}$ & $ 9.1\times 10^{-5}$  & $1.5\times 10^{-4}$ & $1.6\times 10^{-4}$ &  $\sim$ \\
\hline
&&&&&\\
$\Delta m^{2}_{32}(\mu)~({\rm eV^2})$ &  $2.5\times 10^{-3}$ & $2.7\times 10^{-3}$ & $3.3\times 10^{-3}$ & $3.2\times 10^{-3}$ & $\sim$ \\
\end{tabular}
\end{center}
\end{table}


\begin{table}
\caption{Running quark and lepton masses above $M^{}_Z$ in the MSSM
with $\tan\beta=50$, where the matching effect between the MS and
MSSM and the $\overline{\rm MS}$-to-$\overline{\rm DR}$ transition
effect on the input parameters at $M^{}_Z$ have been taken into
account.}
\begin{center}
\begin{tabular}{l|l|l|l|l|l}
& $\mu=M^{}_Z $ & $\mu=1~ {\rm TeV}$ & $\mu=10^9~{\rm GeV}$ & $\mu=10^{12} ~{\rm GeV}$ & $\mu=\Lambda^{}_{\rm GUT}$  \\
\hline
&&&&&\\
$m^{}_u(\mu)~(\rm MeV)$ &  $1.27^{+0.50}_{-0.42}$ & $1.15^{+0.45}_{-0.38}$ &  $0.75^{+0.31}_{-0.26}$ & $0.62^{+0.26}_{-0.22}$ & $0.48^{+0.21}_{-0.17}$ \\
\hline
&&&&&\\
$m^{}_d(\mu)~(\rm MeV)$ & $2.90^{+1.24}_{-1.19}$ & $1.51^{+0.67}_{-0.63}$ & $0.86^{+0.39}_{-0.36}$ & $0.69^{+0.31}_{-0.29}$ & $0.51^{+0.23}_{-0.22}$\\
\hline
&&&&&\\
$m^{}_s(\mu)~(\rm MeV)$ & $55^{+16}_{-15}$ & $29^{+9}_{-8}$ & $16 \pm 5$ & $13 \pm 4$ & $10 \pm 3$ \\
\hline
&&&&&\\
$m^{}_c(\mu)~(\rm GeV)$ & $0.619 \pm 0.084$ & $0.557^{+0.077}_{-0.076} $ & $0.364^{+0.054}_{-0.053}$ & $0.304^{+0.046}_{-0.045}$ & $0.237^{+0.037}_{-0.036}$\\
\hline
&&&&&\\
$m^{}_b(\mu)~(\rm GeV)$ &  $2.89 \pm 0.09$ & $1.54\pm 0.06$ & $0.96 \pm {0.05}$ & $0.79^{+0.05}_{-0.04}$  & $0.61 \pm {0.04}$\\
\hline
&&&&&\\
$m^{}_t(\mu)~(\rm GeV)$ & $171.7 \pm 3.0$  & $161.3 \pm {3.7}$ & $127.0^{+7.4}_{-6.7} $ & $113.2.4^{+8.9}_{-7.7} $  & $94.7^{+10.3}_{-8.4}$\\
\hline
 & $0.486570161$ & $0.286562894$ & $0.254747607$ & $0.235672689$ & $0.206036051$ \\
$m^{}_e(\mu)~(\rm MeV)$ & $\pm {0.000000042}$ & $\pm {0.000000025}$ & $\pm {0.000000022}$ & ${}^{+0.000000021}_{-0.000000020}$ & $\pm {0.000000018}$ \\
\hline
& $102.7181359 $& $60.4961413$ & $53.7834823$ & $49.7577528$ & $43.5020305$\\
$m^{}_\mu(\mu)~(\rm MeV)$ & $\pm 0.0000092 $ & $\pm {0.0000054}$ & $\pm 0.0000048 $ & ${}^{+0.0000045}_{-0.0000044} $ & $\pm 0.0000039$ \\
\hline
&&&&&\\
$m^{}_\tau(\mu)~(\rm MeV)$ &$1746.24^{+0.20}_{-0.19}$ & $1032.61\pm{0.12}$ & $ 937.72 \pm 0.11 $ & $875.31 \pm 0.11$ & $773.44 \pm {0.10}$\\
\hline
\\
{\bf Case A}  \\
\hline
&&&&&\\
$m^{}_1(\mu)~({\rm eV})$ & $0.001$ & $0.001$  & $0.001$  & $0.001$ & $\sim$ \\
\hline
&&&&&\\
$\Delta m^{2}_{21}(\mu)~({\rm eV^2})$  & $8.0\times 10^{-5}$ & $8.7\times 10^{-5}$  & $1.1\times 10^{-4}$ & $1.0\times 10^{-4}$ & $\sim$ \\
\hline
&&&&&\\
$\Delta m^{2}_{32}(\mu)~({\rm eV^2})$  & $2.5\times 10^{-3}$ & $2.7\times 10^{-3}$ & $3.3\times 10^{-3}$ & $3.2\times 10^{-3}$ & $\sim$ \\
\hline
\\
{\bf Case B} \\
\hline
&&&&&\\
$m^{}_1(\mu)~({\rm eV})$ & $0.20$ & $0.21$  & $0.23$  & $0.23$ & $\sim$ \\
\hline
&&&&&\\
$\Delta m^{2}_{21}(\mu)~({\rm eV^2})$ & $8.0\times 10^{-5}$ & $ 1.7\times 10^{-4}$  & $7.1\times 10^{-4}$ & $8.3\times 10^{-4}$ &  $\sim$ \\
\hline
&&&&&\\
$\Delta m^{2}_{32}(\mu)~({\rm eV^2})$ &  $2.5\times 10^{-3}$ & $2.7\times 10^{-3}$ & $3.8\times 10^{-3}$ & $4.1\times 10^{-3}$ & $\sim$ \\
\end{tabular}
\end{center}
\end{table}


\begin{figure}
\vspace{3cm}
\epsfig{file=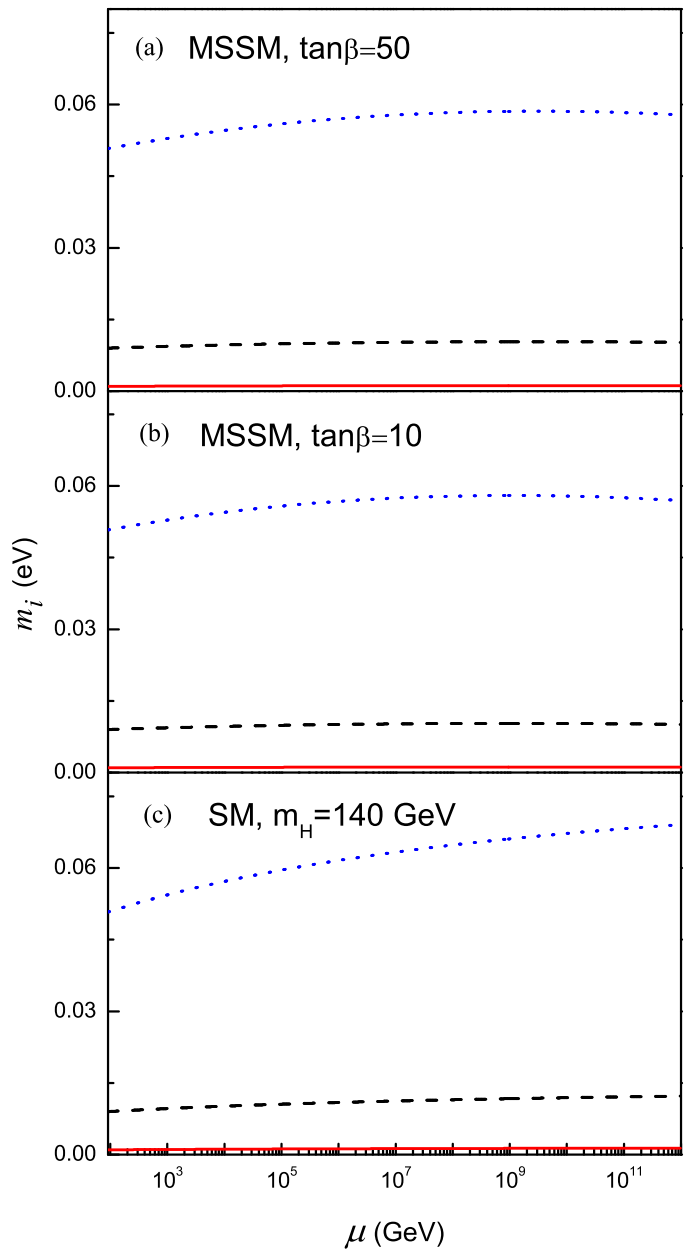,bbllx=-1.5cm,bblly=8cm,bburx=2.5cm,bbury=12cm,%
width=5cm,height=5cm,angle=0,clip=0}\vspace{10cm} \caption{The scale
dependence of three neutrino masses $m^{}_{1}$ (solid line),
$m^{}_{2}$ (dashed line) and $m^{}_{3}$ (dotted line) in the case of
a normal mass hierarchy with $m^{}_1(M^{}_Z)=0.001~ {\rm eV}$ and
$m^{}_1 < m^{}_2 \ll m^{}_3$.}
\end{figure}

\begin{figure}
\vspace{6cm}
\epsfig{file=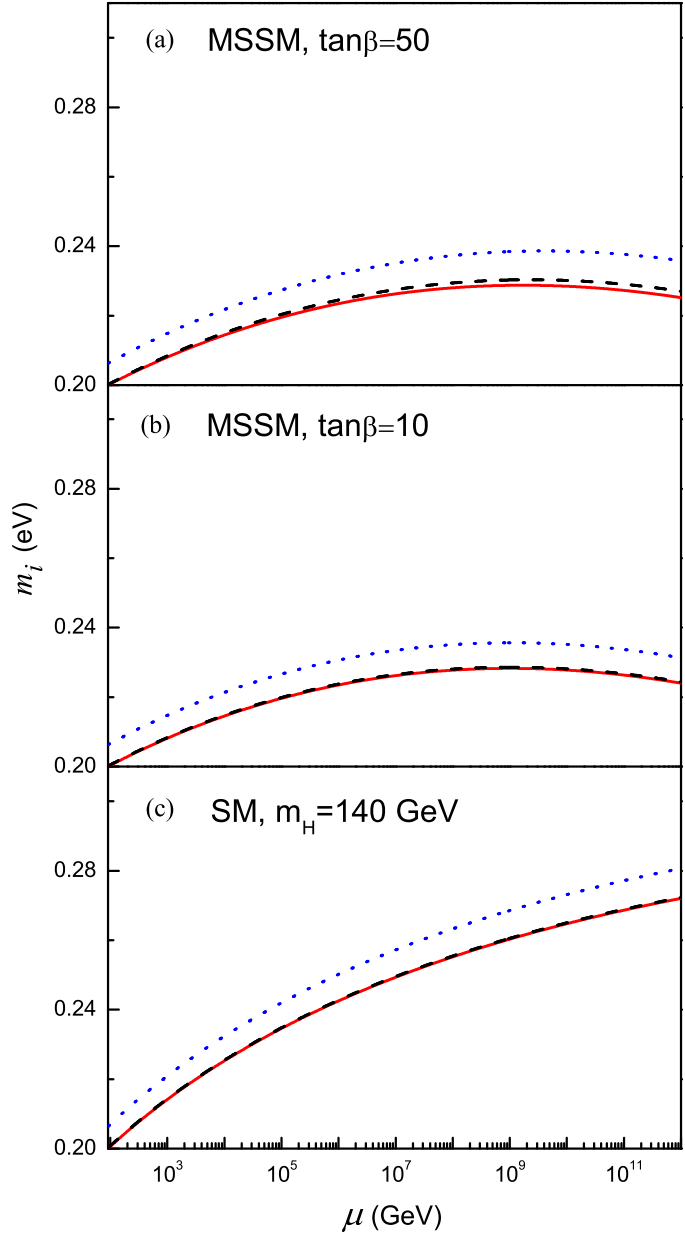,bbllx=-1.5cm,bblly=10cm,bburx=2.5cm,bbury=14cm,%
width=5cm,height=5cm,angle=0,clip=0}\vspace{13cm} \caption{The scale
dependence of three neutrino masses $m^{}_{1}$ (solid line),
$m^{}_{2}$ (dashed line) and $m^{}_{3}$ (dotted line) in the case of
a nearly degenerate mass hierarchy with $m^{}_1(M^{}_Z)=0.2~ {\rm
eV}$ and $m^{}_1 \simeq m^{}_2 \simeq m^{}_3$.}
\end{figure}

\end{document}